\documentstyle[aps,prl,preprint]{revtex}

\begin{document}

\draft

\title{Re-entrant Layer-by-Layer Etching of GaAs(001)}
\author{T.~Kaneko,\cite{byline1} P.~\v{S}milauer,\cite{byline2} and
B.A.~Joyce,\cite{byline3}}
\address{Interdisciplinary Research Centre for Semiconductor Materials,
Imperial College, London SW7~2BZ, United Kingdom}
\author{T.~Kawamura\cite{byline4} and D.D.~Vvedensky\cite{byline5}}
\address{The Blackett Laboratory, Imperial College, London SW7~2BZ,
United Kingdom}

\date{\today}

\maketitle

\begin{abstract}

We report the first observation of re-entrant layer-by-layer etching
based on {\it in situ\/} reflection high-energy electron-diffraction
measurements.  With AsBr$_3$ used to etch GaAs(001), sustained
specular-beam intensity oscillations are seen at high substrate
temperatures, a decaying intensity with no oscillations at intermediate
temperatures, but oscillations reappearing at still lower temperatures.
Simulations of an atomistic model for the etching kinetics reproduce
the temperature ranges of these three regimes and support an
interpretation of the origin of this phenomenon as the site-selectivity
of the etching process combined with activation barriers to interlayer
adatom migration.

\end{abstract}


The development of {\it in situ\/} processing methodologies, where
sample preparation, growth, and post-growth modification are all
carried out in ultra-high vacuum, is crucial for the reproducible
fabrication of atomic-scale heterostructures.  To date, most
fundamental studies of direct relevance to heterostructure formation
have focused on the growth process.  However, the removal of atoms
during processes such as sputtering and etching also provides a way of
manipulating surface properties and morphology.  Several recent studies
have used electron and helium-atom diffraction and scanning tunnelling
microscopy to investigate the atomistic kinetics during the removal of
atoms from a surface by low-energy ion-beam sputtering
\cite{poel,mich1,mich2,bed}. These studies, as well as computer
simulations \cite{chas,sm1,sm2}, have revealed a close correspondence
with the processes that occur during epitaxial growth, with
monovacancies and vacancy islands playing the roles of adatoms and
adatom islands, respectively. It has also been clearly demonstrated by
Tsang {\it et al.\/} \cite{ts1,ts2}, using  reflection high-energy
electron-diffraction (RHEED), that the chemical etching of III-V
semiconductor surfaces by molecular beams of PCl$_3$ and AsCl$_3$ can
occur on a layer-by-layer basis.

The focus of this Letter is on a process related to that studied by
Tsang {\it et al.\/}, the chemical etching of GaAs(001) by AsBr$_3$. By
using {\it in situ\/} RHEED measurements, we observe sustained specular
intensity oscillations at high substrate temperatures, a decaying
intensity with no oscillations at intermediate temperatures, but
oscillations reappearing at still lower temperatures.  To our
knowledge, this is the first report of such re-entrant behavior during
a removal process, although re-entrant layer-by-layer {\it growth\/} was
reported by Kunkel {\it et al.\/} \cite{kunk} for homoepitaxy on
Pt(111). Based on our simulations of the etching kinetics, we propose
that the layer-by-layer removal in the lowest temperature range is due
to a decrease in the vacancy island sizes combined with the
site-selective nature of this etching process.

The experiments were carried out in a conventional molecular-beam
epitaxy (MBE) chamber on nominally singular (0$\pm$0.05$^\circ$)
GaAs(001) substrates. After mounting in the system, the substrates were
pre-cleaned in an As$_2$ flux at $\approx$580$^\circ$C to produce a
well-defined 2$\times$4 reconstructed surface. A buffer layer of
thickness $\approx$1000\AA\ was grown on this surface also at
$\approx$580$^\circ$C from Ga and As$_2$ beams to produce a smooth
starting surface, as indicated by comparatively short, narrow RHEED
streaks. AsBr$_3$ was then introduced directly into the MBE chamber
without a carrier gas.  A concomitant As$_2$ flux was supplied to
avoid any possibility of thermal dissociation of the substrate. The
background pressure during etching was $10^{-8}$ torr. Details of the
experiments will be published elsewhere \cite{kan}.

RHEED measurements were carried out with an electron beam energy of 12.0
keV at an incident angle of $\approx$1$^\circ$ along the [010] azimuth.
Using these diffraction conditions, maxima in the specular beam
intensity correspond to (As-Ga) bilayer increments in the removal of
material for a 2$\times$4 reconstructed surface \cite{shi}.  A slight
complication is the fact that the same surface reconstruction could not
be maintained over the entire temperature range. At high temperatures,
the diffraction pattern showed that the 2$\times$4 reconstruction was
prevalent, while at lower temperatures there appeared to be a mixture
of 2$\times$4 and c(4$\times$4) reconstructions. By analogy with
growth, where a change in reconstruction (usually from 2$\times$4 to
3$\times$1 or the reverse) can produce a change in the ``phase'' or
amplitude of the oscillations \cite{zhang}, but never eliminate them,
we believe that reconstruction changes cannot be responsible for the
re-entrant oscillations observed here.  Only the suppression of the
layer-by-layer growth (etching) mode in which layers are formed
periodically by the nucleation, growth, and coalescence of adatom
(vacancy) islands can eliminate the oscillatory behavior.

The RHEED specular beam intensities are shown in
Fig.~\ref{fig:reentrant} for a range of substrate temperatures.  The
re-entrant behavior of the oscillations is clearly evident. Although
not shown, the temperature range over which the oscillations disappear
is very narrow, extending only over $\approx$10$^\circ$C. The RHEED
pattern in this range shows well-defined spots soon after etching
commences, which is indicative of transmission through
three-dimensional asperities on the surface.  Above and below this
temperature range, the RHEED pattern is streaked over the duration of
the measurement.

The inset to Fig.~\ref{fig:reentrant} shows the substrate
temperature-dependence of the etching rate as deduced from the period
of the RHEED intensity oscillations, calibrated absolutely at
580$^\circ$C by a Talystep measurement on a masked substrate.  This
rate is nearly constant above 420$^\circ$C but diminishes rapidly below
this temperature \cite{reaction}.  We conclude that above 420$^\circ$C
the removal is {\it supply-rate limited\/} and below 420$^\circ$C it is
{\it reaction-rate limited\/}. In particular, the re-entrant behavior
is not due to any pathology in the AsBr$_3$ decomposition rate over the
temperature range of interest, but can be traced to factors involving
surface morphology and the kinetics of the atomistic processes. The
important chemical factor is that the etchant removes Ga atoms as
volatile GaBr$_x$ ($x$$=$$1,2$ or $3$), but our measurements provide no
direct information on reaction pathways.

To model the observations summarized in Fig.~\ref{fig:reentrant}, we
modify a kinetic Monte Carlo model developed for ion-beam sputtering
\cite{sm1,sm2}. We include a site selectivity whereby atoms with low
coordination are preferentially removed from the surface, as described
below. Such site selectivity was investigated previously but found to
be at most of minor importance for ion-beam sputtering of semiconductor
surfaces \cite{chas}. In contrast to the simulations in
Ref.~\cite{chas}, our model leads to an etching rate that is {\it
independent\/} of the surface morphology, as required by the constant
etching rate above 420$^\circ$C shown in Fig.~\ref{fig:reentrant}.

In our simulations, the substrate has a simple cubic structure
\cite{gaas} with neither bulk vacancies nor overhangs allowed (the
solid-on-solid model). Atoms are removed from the substrate at the
experimentally observed rate of etching by first selecting a site
randomly, then searching within a square of a fixed linear size $L$,
centered at the originally selected site, for the surface atom with
the fewest lateral nearest neighbors. This atom is then removed from
the surface. Surface migration is modeled as a nearest-neighbor hopping
process with the rate $k(E,T)$$=$$k_{0}\exp(-E/k_{B}T)$, where $E$ is
the hopping barrier, $T$ is the substrate temperature, and $k_B$ is
Boltzmann's constant. The pre-factor $k_{0}$ is taken to be the
vibrational frequency of  a surface adatom and assigned the value
$2k_{B}T/h$ where $h$ is Planck's constant. The hopping barrier is the
sum of a substrate term, $E_{S}$, a lateral nearest-neighbor
contribution, $E_{NN}$, and a step-edge barrier. The nearest-neighbor
contribution is $E_{NN}$$=$${1\over2}n_{1}E_{N}$$+$$E^{\prime}_{N}$,
where $E^{\prime}_{N}$ is nonzero only if the number of lateral nearest
neighbors before a hop, $n_{1}$, is greater than that after the hop,
$n_{2}$, in which case $E^{\prime}_{N}$$=$${1\over2}(n_{1}-n_{2})E_{N}$.
This definition increases diffusion along island edges and leads to
high vacancy mobility (consistent with experimental observations
\cite{poel,mich1,mich2,bed}) but preserves the difference in the
hopping rate of edge atoms with different coordination as well as the
activation barriers to detachment of atoms from step edges
($E_{NN}\!=\!n_1E_N$ for $n_2\!=\!0$). The vicinity of a step is
detected by counting the number of next-nearest neighbors in layers
above and below the hopping atom before ($m_1$) and after ($m_2$) a
hop. An additional (step-edge) barrier to hopping is present only if
$m_1$$>$$m_2$, in which case it equals $(m_1$$-$$m_2)E_B$, where
$E_B$ is a model parameter.

The simulations were carried out on 200$\times$200 lattices with
periodic boundary conditions and $E_{S}$$=$$1.58$~eV,
$E_{N}$$=$$0.24$~eV, $E_B$$=$$0.175$~eV. When used for simulations of
the {\it growth\/} of GaAs(001), these parameters produced excellent
quantitative agreement with RHEED measurements for a range of growth
conditions \cite{shi,sm3}. The parameter $L$ was set equal to 3.
Comparisons with the RHEED measurements are based on the continuous
monitoring of the surface step density, which is an approximate analog
of the RHEED specular-beam intensity under the diffraction conditions
used \cite{shi,sm3}. We also calculate the kinematic intensity at
anti-Bragg diffraction conditions, which is relevant for other
diffraction techniques (e.g., He-atom scattering, grazing-incidence
X-ray diffraction). All results presented are averaged over at least
five independent simulations.

The results of the simulation are shown in Fig.~\ref{fig:simulation}.
The re-entrant nature of the RHEED oscillations is seen to be
reproduced by the behavior of the step density in the simulations.
Kinematic intensity oscillations do not disappear in the intermediate
temperature, but are much better defined for both lower and higher
temperatures.  Notice that the ranges of temperatures at which the
oscillations are observed correspond well to those in
Fig.~\ref{fig:reentrant}. The temperature range over which the
oscillations disappear (not shown) is very narrow, in qualitative
agreement with the experiment.

Based on the experimental measurements and our simulations, the
explanation of the re-entrant etching oscillations we propose is as
follows. Etching creates surface vacancies which diffuse on the surface
and form islands. As these islands grow, new vacancies are created in
lower layers within the flat regions near their centers. At high
temperatures ($\gg\!460^\circ$C), there are many free adatoms on the
surface. Due to the high thermal mobility of adatoms and vacancies,
the vacancies created in lower layers are filled in \cite{ann}.
This results in the periodic removal of individual layers through the
formation, accretion, and coalescence of vacancy islands.  The
interlayer atomic transport in this temperature range is appreciable
because the high density of free adatoms dominates the inhibiting
effect of the step--edge barrier.

At intermediate temperatures ($\approx\!460^\circ$C) the number of free
adatoms on the surface is reduced drastically because the rate of their
detachment from step edges decreases exponentially with decreasing
temperature. The filling in of vacancies created in lower layers is now
strongly hindered by the step-edge barrier and, therefore, vacancy
islands in lower layers are formed. These islands are nested inside
vacancy islands on higher layers, creating pits (inverse pyramids) in
the surface (cf. the scanning-tunneling microscopy images in
Refs.~\cite{mich1} and \cite{mich2}).  As a result of this {\it
multilayer\/}, or {\it three-dimensional\/}, mode of removal, a rough
surface morphology quickly develops which is qualitatively different
from both the low-- and high--temperature morphologies (see
Refs.~\cite{mich3} and \cite{sm4} for the growth analog) and is
accompanied by the disappearance of RHEED oscillations and the
appearance of a spotty diffraction pattern.

Finally, at low temperatures ($\ll\!460^\circ$C), the mobility of
vacancies is low, which leads to the formation of many small islands
\cite{vill} and a high density of atoms with low coordination in the
uppermost layer.  These atoms are preferentially removed by the etching
process which leads to the re-emergence of an approximately
layer-by-layer removal mode and the reappearance of RHEED
oscillations.  The transition from the layer-by-layer removal regime at
high temperatures to the multilayer removal regime has been observed in
the ion-beam sputtering of Pt(111) \cite{poel,mich1,mich2}, but
obtaining the reappearance of the oscillations at lower temperatures
requires a site-selective step such as the one we have proposed
\cite{AsBr3}.

There is an important consequence of our interpretation of the
reappearance of RHEED oscillations at low temperatures. If the etching
rate is increased in the intermediate temperature regime by supplying
more etchant molecules to the substrate, the oscillations
should reappear, as if the substrate temperature had been
decreased with the original flux maintained. In both cases the vacancy
islands become smaller \cite{vill}, the density of low-coordinated
atoms at the uppermost layer is higher and the site-selective step
becomes more important. Fig.~\ref{fig:flux} demonstrates that this is
indeed the case:~as the etching rate is increased, oscillations
reappear at 460$^\circ$C (cf. Fig.~\ref{fig:reentrant}) and their
amplitude grows with the etching rate. An analogous effect is observed
in the simulations. Note that the oscillations reappear even for a
modest change of the etching rate (a factor of 2) due to the vacancy
islands at the chosen temperature being just large enough to make the
preferential etching of atoms with low coordination unimportant. Thus,
even a small decrease in the island sizes brings the oscillations back.
The result shown in Fig.~\ref{fig:flux} demonstrates once more that the
re-entrance of oscillations involves the surface morphology and not the
change in the etching reaction rate with the temperature.

In conclusion, we have reported the first observation of the re-entrant
layer-by-layer etching of a solid surface. Based on simulations of
the etching kinetics, we propose that the RHEED oscillations indicating
layer-by-layer etching disappear in the intermediate temperature regime
because of the increasing importance of the barriers to interlayer
hopping and the decreasing number of free adatoms on the surface.  The
reappearance of the oscillations at still lower temperatures is due to
a decrease of the vacancy island sizes combined with the site-selective
nature of the etching process whereby atoms with low coordination are
preferentially removed from the surface.

We acknowledge gratefully the support of Imperial College and the
Research Development Corporation of Japan under the auspices of the
``Atomic Arrangement:~Design and Control for New Materials'' Joint
Research Program.

\begin{figure}
\caption{RHEED specular-beam intensity evolution during etching of
GaAs(001) with an AsBr$_3$ flux of 0.36 sccm and different substrate
temperatures. The inset shows the etching rate vs.~the substrate
temperature.}
\label{fig:reentrant}
\end{figure}

\begin{figure}
\caption{The surface step density and the specular--beam intensity
calculated in the kinematic approximation for an anti-Bragg angle of
incidence during growth at three different substrate temperatures. The
step density increases {\it downwards\/}. The upper two curves in both
plots are given offsets to make comparisons easier.}
\label{fig:simulation}
\end{figure}

\begin{figure}
\caption{RHEED specular-beam intensity evolution during etching of
GaAs(001) at a substrate temperature of 460$^\circ$C but with different
AsBr$_3$ fluxes.}
\label{fig:flux}
\end{figure}


\begin{references}
\bibitem{reaction} The oscillation period below 420$^\circ$C decreases
as $\exp(-E_A/k_BT)$, $E_A\approx 1.8$~eV, which can be modeled with an
effective activation energy for the removal process. Because our main
focus is on re-entrant layer-by-layer etching, we have not included
this step in our model.
\bibitem{gaas} We assume that the removal of Ga atoms is the
rate-limiting step and that the As kinetics and the effects of surface
reconstructions can be incorporated into the effective model parameters
\cite{shi,sm3}.
\bibitem{ann} Possible mechanisms for the annihilation of vacancies
formed in lower layers and the role of adatom and vacancy migration are
discussed in Refs. \cite{mich2} and \cite{sm1}.
\end{references}
\end{document}